\documentclass[aps, showpacs, showkeys, nofootinbib, floatfix, superscriptaddress]{revtex4}

\usepackage{amsfonts}
\usepackage{amssymb}
\usepackage{amsmath}
\usepackage{graphicx}
\usepackage{float}
\usepackage{epstopdf}
\usepackage{soul}
\usepackage{makecell}
\usepackage{hyperref}
\usepackage{multirow}  
\usepackage{subcaption}

\begin{document}

\title{Photon Motion and Shadows of Rotating Black Holes with Nonlinear Electromagnetic and Anisotropic Matter Fields}
\author{Mou Xu}
\affiliation{Department of Physics, Liaoning Normal University, Dalian 116029, P. R. China}
\author{Jianbo Lu}
\email{lvjianbo819@163.com}
\affiliation{Department of Physics, Liaoning Normal University, Dalian 116029, P. R. China}
\author{Yu Liu}
\affiliation{Department of Physics, Liaoning Normal University, Dalian 116029, P. R. China}
\author{Shumin Wu}
\affiliation{Department of Physics, Liaoning Normal University, Dalian 116029, P. R. China}

\begin{abstract}

This paper investigates the effects of the nonlinear electromagnetic field and the anisotropic matter field on photon motion, shadow structures, and the energy emission rate of a rotating black hole (BH). Using the Hamilton-Jacobi formalism, we derive the photon motion equations and analyze the distribution and stability of photon regions. The results show that the anisotropic matter field parameters affect the size and shape of the photon region outside the event horizon more significantly than the nonlinear electromagnetic field parameter. As the anisotropic matter field parameter $K$ decreases, the unstable photon region outside the BH gradually expands and becomes increasingly flattened. Furthermore, we construct the BH shadow in terms of the celestial coordinates and obtain the corresponding shadow images by backward ray tracing. Several shadow observables, including the shadow radius, distortion parameter, shadow area, and oblateness, are also analyzed. The results indicate that the anisotropic matter field affects the shadow size more strongly than the shadow shape. Specifically, the shadow radius and area both decrease significantly as the parameter $K$ decreases, but increase markedly as the anisotropic matter state parameter $\omega$ increases. In addition, we analyze the energy emission rate of the BH and find that decreasing $K$ or increasing magnetic charge $Q$ suppresses its peak value, while the influence of $\omega$ remains comparatively mild. These results provide a useful reference for understanding the effects of nonlinear electromagnetic and anisotropic matter fields on rotating BH shadows and related observational signatures.
\end{abstract}

\maketitle

\section{$\text{Introduction}$}

In realistic astrophysical environments, BHs are generally surrounded by accretion matter, electromagnetic fields, dark matter halos, or other effective matter fields \cite{Karthik2025Perturbations,Lee2021Shadow,Cho2019Fluid}. Isotropic matter distributions have been extensively discussed in studies of BHs and compact objects \cite{Cho2017Static,Katsuragawa2016Relativistic,Hendi2017Neutron,Hendi2018Holographical}. On the other hand, for certain ranges of matter density, the pressures of the matter field along different directions may exhibit distinct properties, thereby giving rise to local anisotropy \cite{Herrera1997Local}. Such anisotropy may further affect the structure and physical properties of self-gravitating systems. Consequently, the study of anisotropic matter fields has attracted increasing attention \cite{Bowers1974Anisotropic,Mak2003Anisotropic,Dev2002Anisotropic}. In particular, for certain static and spherically symmetric BHs, an anisotropic matter field with the radial pressure equal to the negative energy density can make the energy density and pressure continuous at the event horizon. This provides a possible condition for a static configuration involving a BH and an anisotropic matter field \cite{Kim2020Rotating,Cho2019Simple}. The interaction between a BH and its surrounding matter fields can alter the spacetime structure in its vicinity and further affect its observable characteristics. Therefore, studying BHs in the presence of anisotropic matter fields is important for understanding strong gravity effects and their observational signatures.

Nonlinear electrodynamics is a nonlinear generalization of Maxwell's electromagnetic theory \cite{Sorokin2022Introductory} and has been widely applied in BH physics. Bardeen first proposed a regular BH model \cite{Bardeen1968Nonsingular}. Ay{\'o}n-Beato and Garc{\'i}a later showed that this model can be interpreted as a nonlinear magnetic monopole solution arising from the coupling of Einstein gravity to nonlinear electrodynamics \cite{AyonBeato2000TheBardeen}. Within this framework, various BH solutions have been constructed by introducing nonlinear electromagnetic fields into gravitational theories, including the Ay{\'o}n-Beato-Garc{\'i}a BH, the Hayward BH, and the Berej-Matyjasek-Tryniecki-Wornowicz BH \cite{AyonBeato1998Regular,Hayward2006Formation,Berej2006Regular}. Recently, a static and spherically symmetric BH solution was obtained by simultaneously incorporating a nonlinear electromagnetic field and an anisotropic matter field. The corresponding rotating solution was further constructed through a modified Newman-Janis algorithm and is referred to as the anisotropic nonlinear magnetic charged rotating black hole (ANRBH) \cite{Li2025Rotating}.

Black holes, as one of the most important predictions of general relativity, have long been regarded as key objects in theoretical and observational studies of the strong gravity regime. Since the release of the M87* and Sgr A* images by the Event Horizon Telescope (EHT) Collaboration \cite{Azulay2019FirstM87,Akiyama2019FirstM87,Akiyama2022FirstSagittariusA}, BH shadows have become an important observational probe for studying light propagation in strong gravitational fields and the spacetime structure of BHs \cite{Zeng2025Optical,Xu2025Particle,Abdujabbarov2016Shadow,Ling2022Theshadows,Ban2025Investigating,Guo2024Image}. Early studies of BH shadows can be traced back to the analysis of light propagation in the Schwarzschild spacetime by Synge \cite{Synge1966Theescape}. Subsequently, Bardeen investigated the shadow structure of the Kerr BH \cite{Bardeen1973BlackHoles}. In general, the shadow of a static and spherically symmetric BH is circular. For a rotating BH, however, the spin parameter and the observer's inclination angle can affect the shadow, which can exhibit displacement and distortion \cite{Bardeen1973BlackHoles,Hioki2009Measurement}. The formation of a BH shadow is closely related to unstable photon orbits. 
In spherically symmetric spacetimes, these orbits correspond to the photon sphere, whereas in rotating BH spacetimes they are generally no longer confined to a single spherical surface but form a photon region \cite{Volker2004Gravitational,Falcke2000Viewing,Gralla2019Blackholeshadows,Chen2023Blackholeimages}. Accordingly, BH shadows have been widely used to investigate the geometrical properties and potential observational signatures of different BH spacetimes \cite{Wang2023Rings,Liu2025Charged,Xiong2025Investigating,Hu2024Influences,Liu2024KerrMOG,Xu2025Optical}.

External matter fields can alter the spacetime structure around BHs, thereby affecting photon motion, photon regions and shadow structures \cite{Fernandes2025Spinning,Fathi2024Blackhole,Kurmanov2026Effects}. Anisotropic matter fields usually introduce additional physical parameters into BH spacetimes, and further influence BH properties and their observational signatures \cite{Lee2021Shadow,Kim2020Rotating,Badia2020Influence}. In recent years, the influence of matter fields around BHs on photon motion and shadow structures has received considerable attention \cite{Gao2023Investigating,Li2025Images,Guo2023Shadow,Wang2022Shadow}. Previous studies have shown that anisotropic matter fields can affect observable phenomena such as gravitational wave signals and BH shadows \cite{Karthik2025Perturbations,Bolokhov2022On}. In static and spherically symmetric BH spacetimes, an anisotropic matter field can alter the optical appearance of the BH shadow \cite{Kurmanov2026Effects}. Furthermore, the shadows of Kerr-Newman-like BHs in the presence of anisotropic matter fields have also been investigated \cite{Badia2020Influence}. In the framework of Rastall gravity, the anisotropic fluid parameter and the Rastall coupling parameter can reduce the size of the rotating BH shadow relative to the Kerr case \cite{Kumar2021Shadows}. Beyond shadow studies, recent works have also analyzed the effects of anisotropic matter fields on particle collisions and considered wormhole solutions involving anisotropic matter \cite{Ahmed2021Rotating,Kim2019Spherically}. Based on these studies, this work further investigates a rotating BH with both a nonlinear electromagnetic field and an anisotropic matter field, focusing on its photon motion and shadow properties.

This paper is organized as follows. In Sec. II, we briefly introduce the anisotropic nonlinear magnetic charged rotating BH solution. Sec. III derives the equations of photon motion in this BH spacetime within the Hamilton-Jacobi formalism. The conditions for spherical photon orbits and the photon region are then presented. With these conditions, we analyze the effects of the nonlinear electromagnetic field and the anisotropic matter field on photon motion. Based on the celestial coordinates, we construct the BH shadow and discuss the dependence of the shadow on the spin, magnetic charge, and anisotropic matter field parameters in Sec. IV. Furthermore, we calculate several shadow observables, such as the shadow radius and shadow area. In addition, we discuss the energy emission rate of this BH and analyze its dependence on different physical parameters. Finally, Sec. V summarizes our results. Throughout this paper, we adopt geometrized units and set $G=c=M=1$, with the metric signature $(-,+,+,+)$.

\section{$\text{Anisotropic Nonlinear Magnetic Charged Rotating Black Hole}$}

In realistic astrophysical environments, BHs are usually not isolated, but may be influenced by surrounding matter fields and other environmental factors. Owing to strong gravitational effects near the event horizon, ordinary matter is generally difficult to maintain in a static configuration with a BH \cite{Cho2019Simple}. By contrast, anisotropic matter fields with negative radial pressure can be used to study static configurations involving BHs and matter fields \cite{Kim2020Rotating,Cho2019Simple}. Meanwhile, nonlinear electrodynamics generalizes Maxwell's theory and can describe electromagnetic field effects beyond Maxwell theory in strong field regions \cite{Sorokin2022Introductory}. Motivated by these considerations, a theoretical framework incorporating both nonlinear electromagnetic fields and anisotropic matter fields has been developed, and the corresponding BH solutions have been constructed \cite{Li2025Rotating}.

Within the framework of Einstein gravity with a nonlinear electromagnetic field and an anisotropic matter field, the corresponding field equations can be written as \cite{Li2025Rotating}:
\begin{equation}
G_{\mu}^{\nu}=2\left[\frac{\partial \mathcal{L}(F)}{\partial F} F_{\mu \rho} F^{\nu \rho}-\delta_{\mu}^{\nu} \mathcal{L}(F)\right] +8\pi  T_{\mu}^{\nu},
\end{equation}
\begin{equation}
\nabla_{\mu}\left(\frac{\partial \mathcal{L}(F)}{\partial F} F^{\mu \nu}\right)=0 ,
\end{equation}
\begin{equation}
\nabla_{\mu} {}^\ast F^{\mu\nu}=0,
\end{equation}
where ${}^\ast F^{\mu\nu}$ denotes the dual tensor of the electromagnetic field strength tensor $F_{\mu \nu}$ \cite{DeLorenci2001Dyadosphere}. The function $\mathcal{L}(F)$ is the Lagrangian density of the nonlinear electromagnetic field and depends on the electromagnetic invariant $F$:
\begin{equation}
F \equiv \frac{1}{4} F_{\mu \nu} F^{\mu \nu},
\end{equation}
here, the electromagnetic field strength tensor $F_{\mu \nu}$ is defined in terms of the electromagnetic four-potential $A_{\mu}$ as:
\begin{equation}
F_{\mu \nu}=\partial_{\mu} A_{\nu}-\partial_{\nu} A_{\mu}.
\end{equation}
The Lagrangian density of the nonlinear electromagnetic field is given by \cite{Nam2018On,Benavides2020Rotating}:
\begin{equation}
\mathcal{L}(F)=\frac{3 M}{|Q|^{3}} \frac{\left(2 Q^{2} F\right)^{3 / 2}}{\left[1+\left(2 Q^{2} F\right)^{3 / 4}\right]^{2}},
\end{equation}
where $Q$ is the magnetic charge parameter associated with the nonlinear electromagnetic field \cite{Nam2018On}. For an anisotropic fluid, its energy-momentum tensor can be written as \cite{Kim2020Rotating,Cho2019Simple}:
\begin{equation}
T_{\nu}^{\mu}=\operatorname{diag}\left(-\varepsilon, p_{r}, p_{\theta}, p_{\phi}\right).
\end{equation}
Furthermore, the radial and tangential pressures are chosen as $p_{r}=-\varepsilon(r)$ and $p_{\theta}=p_{\phi}=\omega \varepsilon(r)$, respectively. The corresponding energy density $\varepsilon(r)$ can be expressed as \cite{Cho2019Simple}:
\begin{equation}
\varepsilon(r)=\frac{(1-2\omega)K}{8\pi r^{2\omega+2}},
\end{equation}
where $K$ is a constant related to the anisotropic matter field. From the condition $\varepsilon(r)\ge0$, one finds that the anisotropic matter field parameters should satisfy $(1-2\omega)K \ge 0$. Within the above theoretical framework, the spherically symmetric BH solution incorporating the nonlinear electromagnetic field and the anisotropic matter field can be obtained as \cite{Li2025Rotating}:
\begin{equation}
d s^{2}=-(1-\frac{2 M r^{2}}{r^{3}+Q^{3}}-\frac{K}{r^{2\omega }}) d t^{2}+(1-\frac{2 M r^{2}}{r^{3}+Q^{3}}-\frac{K}{r^{2\omega }})^{-1} d r^{2}+r^{2} (d \theta^{2}+\sin ^{2} \theta d \phi^{2}),	
\label{solution1}	
\end{equation}
where $M$ denotes the BH mass. The anisotropic matter field in this BH solution is characterized by the parameters $\omega$ and $K$. Here, $\omega$ is the state parameter of the anisotropic matter field, while $K$ is a global charge associated with the anisotropic matter field. When $K=0$, this solution reduces to the Hayward-like BH \cite{Hayward2006Formation}. When $Q=K=0$, Eq. (\ref{solution1}) further reduces to the Schwarzschild BH. It has been found that, under the influence of the anisotropic matter field, the curvature invariants diverge at $r=0$. Therefore, the anisotropic nonlinear magnetic charged BH is not a regular BH \cite{Li2025Rotating}.

The spherically symmetric BH solution in Eq. (\ref{solution1}) can be extended to an axisymmetric rotating BH solution by applying the modified Newman-Janis algorithm \cite{Li2025Rotating}:
\begin{equation}
\begin{gathered}
d s^{2}=  -\left[1-\frac{2 \rho r}{\Sigma}\right] d t^{2}+\frac{\Sigma}{\Delta} d r^{2}-\frac{4 a \rho r \sin ^{2} \theta}{\Sigma} d t d \phi +\Sigma d \theta^{2}+\sin ^{2} \theta\left[r^{2}+a^{2}+\frac{2 a^{2} \rho r \sin ^{2} \theta}{\Sigma}\right] d \phi^{2},\\
\Delta =r^{2}-2 \rho r+a^{2}, \\
\Sigma =r^{2}+a^{2} \cos ^{2} \theta, \\
2 \rho =\frac{2 M r^{3}}{r^{3}+Q^{3}}+K r^{-2\omega  +1},
\end{gathered}
\label{solution2}
\end{equation}
where $a$ is the spin parameter of the BH. For $a=0$, this metric reduces to the spherically symmetric form in Eq. (\ref{solution1}). When $Q=K=0$, Eq. (\ref{solution2}) further reduces to the Kerr BH. Previous studies have shown that the spacetime is asymptotically flat for $\omega>1/2$ \cite{Cho2019Simple}. Therefore, in this paper we focus on the case $\omega>1/2$. Combined with the non-negative energy density condition, this implies $K \leq 0$. For convenience in the subsequent numerical calculations, we introduce the following dimensionless parameters:
\begin{equation}
r^*=\frac{r}{M}, \quad a^*=\frac{a}{M}, \quad Q^*=\frac{Q}{M}, \quad K^*=\frac{K}{M^{2\omega}}.
\end{equation}
For notational simplicity, the asterisks on the dimensionless quantities will be omitted hereafter. The BH shadow is determined by photon motion in the strong field region. We therefore turn to null geodesics in this spacetime.

\section{$\text{Photon Regions around the Anisotropic Nonlinear Magnetic Charged Rotating Black Hole}$}

To study how the nonlinear electromagnetic field and the anisotropic matter field affect the BH shadow, we first analyze photon motion in the corresponding background spacetime. A BH shadow is essentially determined by the critical boundary between light rays escaping to infinity and those captured by the BH, and this boundary is generally closely related to unstable spherical photon orbits in the vicinity of the BH \cite{Teo2003Spherical,Bronzwaer2021Thenature}. In the Schwarzschild spacetime, the unstable circular photon orbits are located at $r=3M$. Owing to spherical symmetry, these orbits constitute the photon sphere \cite{Teo2003Spherical,Shoom2017Metamorphoses}. For Kerr BHs, by contrast, the photon sphere is replaced by a photon region containing spherical null geodesics \cite{Grenzebach2014Photon,Grenzebach2015Photon}. The photon region is determined by the background geometry and encodes important information about photon motion near the BH. Motivated by this, we study the equations of photon motion and the photon region of the rotating BH considered in this work. This analysis helps clarify the dynamical behavior of null geodesics near the BH and provides the basis for the subsequent discussion of the shadow shape and its dependence on the physical parameters.

Next, we derive the photon equations of motion around the ANRBH from the Hamilton-Jacobi (H-J) formalism. For photons, the H-J equation reads \cite{Perlick2002Fermat}:
\begin{equation}
\frac{\partial S}{\partial \lambda}+H= 0,
\label{HJ}
\end{equation}
where $S$ is the Hamilton-Jacobi action, and $\lambda$ is the affine parameter along the null geodesic. The Hamiltonian for photon motion is given by:
\begin{equation}
H= \frac{1}{2} g^{\mu\nu} \frac{\partial S}{\partial x^{\mu}} \frac{\partial S}{\partial x^{\nu}}=0.
\end{equation}
Therefore, the H-J equation for photons in Eq. (\ref{HJ}) can be rewritten as:
\begin{equation}
\frac{\partial S}{\partial \lambda}+\frac{1}{2} g^{\mu\nu} \frac{\partial S}{\partial x^{\mu}} \frac{\partial S}{\partial x^{\nu}}=0.
\end{equation}
For the stationary and axisymmetric spacetime described by Eq. (\ref{solution2}), the metric components are independent of the coordinates $t$ and $\phi$. Consequently, photon motion admits two conserved quantities: the energy $E$ and the axial angular momentum $L_z$. Following the Carter separation method, we adopt the following separable ansatz for the Hamilton-Jacobi action \cite{Chandrasekhar1998Themathematical}:
\begin{equation}
S = -E t + L_z \phi + S_r(r) + S_\theta(\theta).
\label{carter}
\end{equation}

To obtain the first-order equations of motion for photons around the ANRBH, we use the covariant four-momentum defined by:
\begin{equation}
p_{\mu}=\frac{\partial S}{\partial x^{\mu}}=g_{\mu \nu}\frac{d x^{\nu}}{d \lambda}.
\end{equation}
Combining this relation with the separable ansatz in Eq. (\ref{carter}) gives:
\begin{equation}
p_t= \frac{\partial S}{\partial t}=-E ,
\end{equation}
\begin{equation}
p_{\phi}=\frac{\partial S}{\partial \phi}=L _z.
\end{equation}
The energy $E$ and angular momentum $L_z$ can be obtained as:
\begin{equation}
E =-g_{tt}\dot{t}-g_{t \phi }\dot{\phi},
\end{equation}
\begin{equation}
L _z= g_{t \phi}\dot{t}+g_{\phi \phi}\dot{\phi},
\end{equation}
where the dot denotes differentiation with respect to the affine parameter $\lambda$. The equations of motion for photons around the ANRBH are given by:
\begin{equation}
\Sigma \dot{t}=\frac{E\sin^{2} \theta[(r^{2}+a^{2})^2-a^2\Delta \sin^{2} \theta]\Sigma^2-2\rho r a L_z\sin^{2} \theta \Sigma^2}{(2\rho ra\sin^{2} \theta)^2+[(r^{2}+a^{2})^2-a^2\Delta \sin^{2} \theta]\sin^{2} \theta(\Sigma-2\rho r)},
\label{motiont}
\end{equation}
\begin{equation}
\Sigma \dot{r} = \sqrt{[E(r^{2}+a^{2})-aL_z]^2-\Delta[\mathcal{K}+(L_z-aE)^2]}\equiv \sqrt{\mathcal{R}(r)},
\label{motionr}
\end{equation}
\begin{equation}
\Sigma \dot{\theta} = \sqrt{\mathcal{K}+a^2E^2\cos^2\theta-L_z^2\cot^2\theta}\equiv \sqrt{\Theta(\theta)},
\label{motiontheta}
\end{equation}
\begin{figure}[H]
\centering
\includegraphics[width=0.24\textwidth]{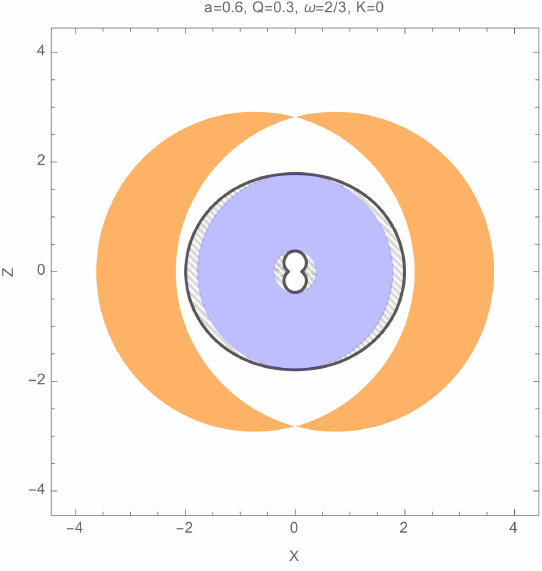}
\includegraphics[width=0.24\textwidth]{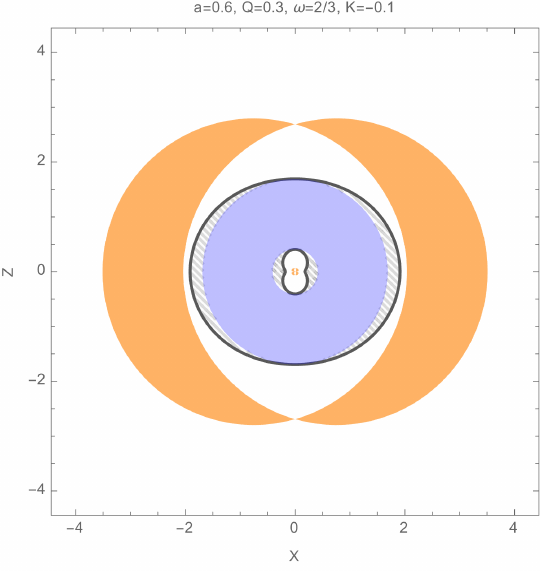}
\includegraphics[width=0.24\textwidth]{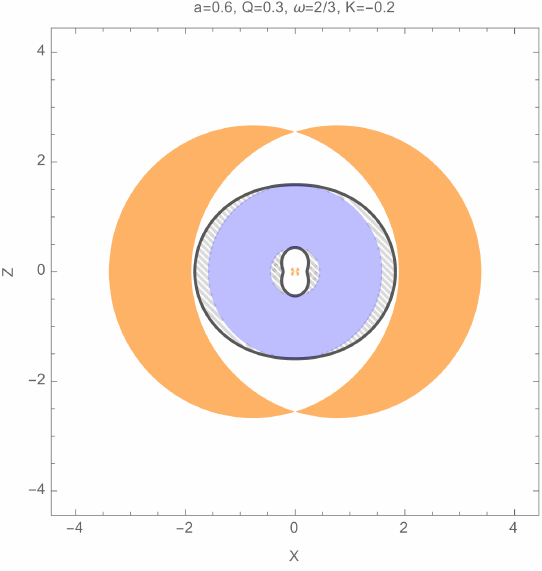}
\includegraphics[width=0.24\textwidth]{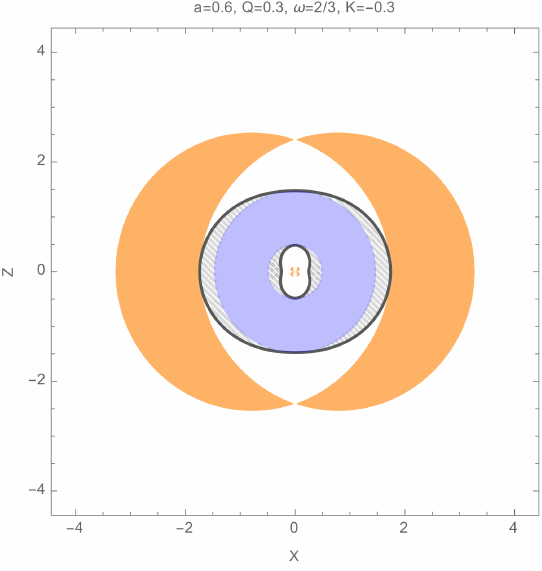}
\includegraphics[width=0.24\textwidth]{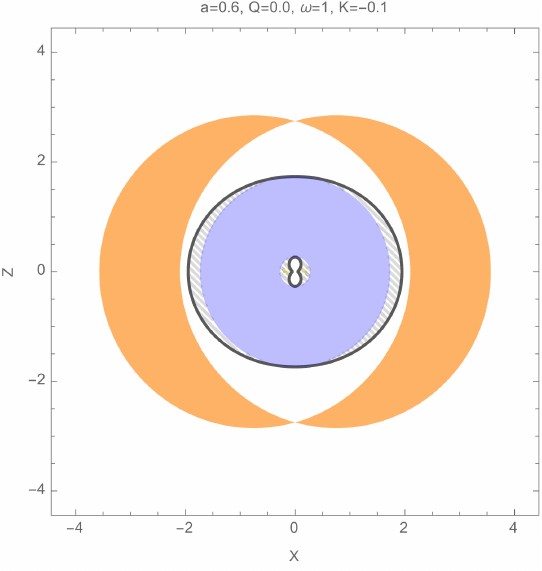}
\includegraphics[width=0.24\textwidth]{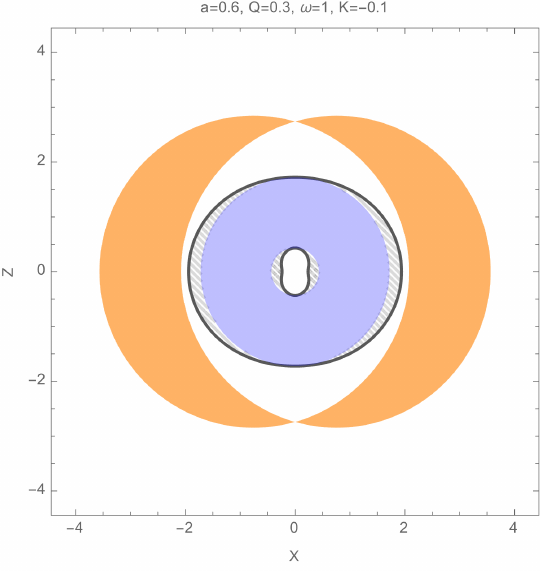}
\includegraphics[width=0.24\textwidth]{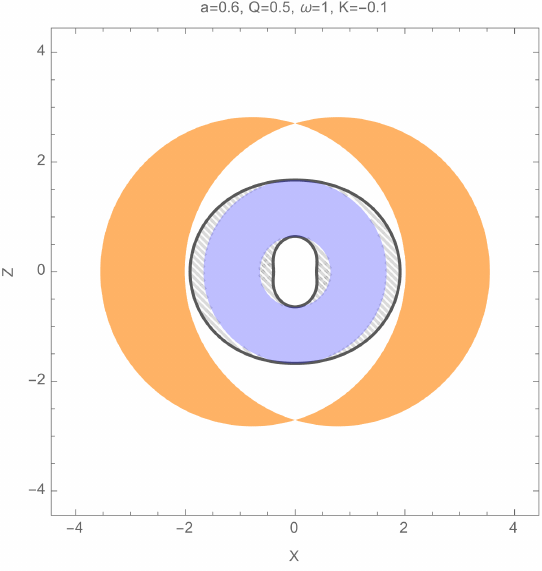}
\includegraphics[width=0.24\textwidth]{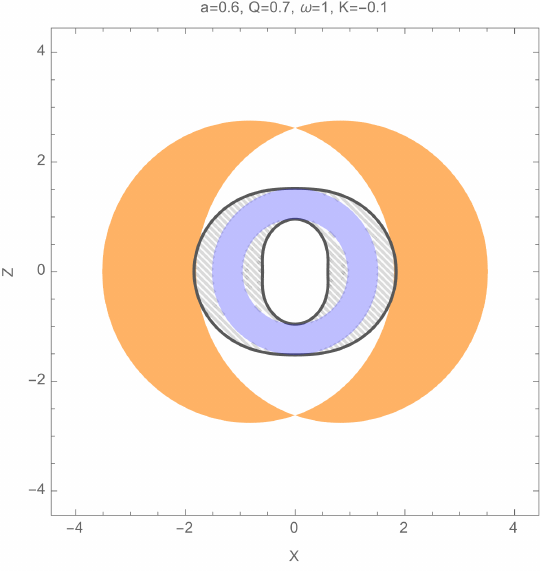}
\includegraphics[width=0.24\textwidth]{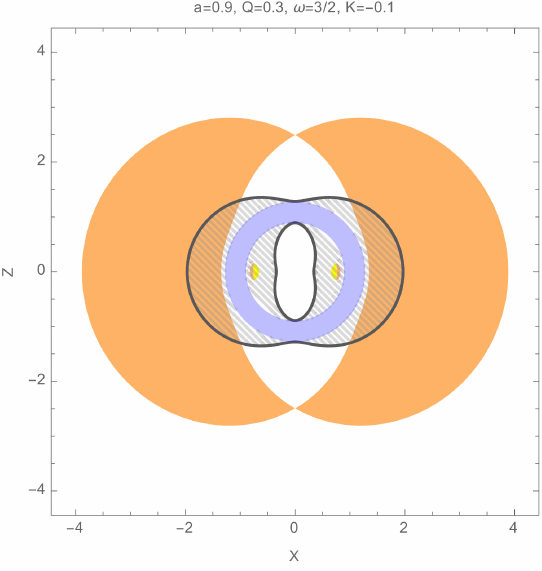}
\includegraphics[width=0.24\textwidth]{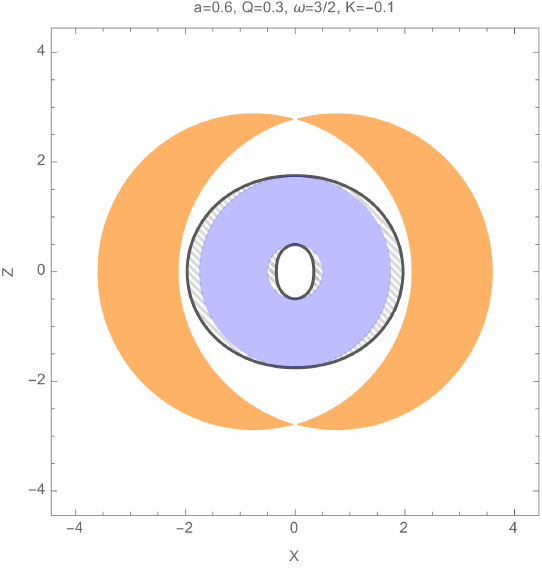}
\includegraphics[width=0.24\textwidth]{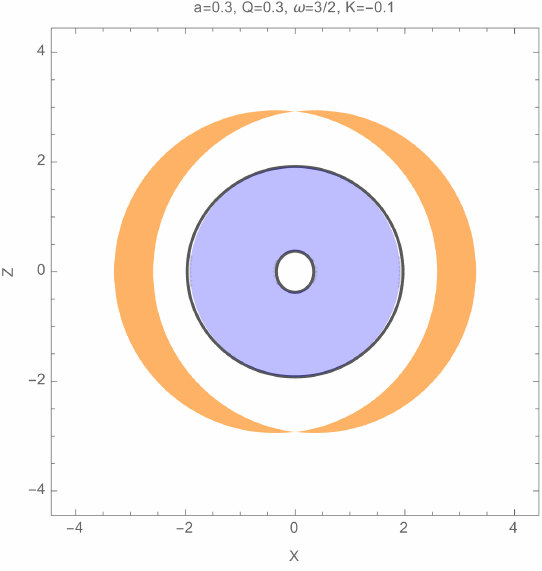}
\includegraphics[width=0.24\textwidth]{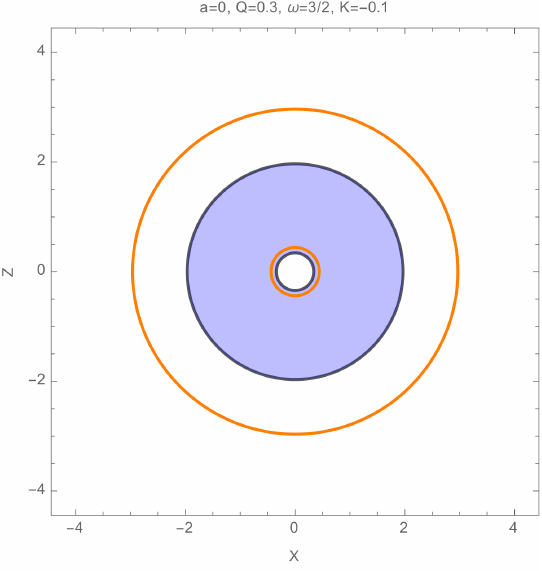}
\caption{Photon regions of the anisotropic nonlinear magnetic charged rotating black hole in the ($xz$) plane. The boundaries of the blue regions represent the black hole horizons, the gray hatched regions denote the ergoregion structure, and the orange and yellow regions correspond to unstable and stable photon regions, respectively.}
\label{figphoton}
\end{figure}
\begin{equation}
\Sigma \dot{\phi}=\frac{2\rho raE\sin^{2} \theta \Sigma^2+L_z(\Sigma-2\rho r)\Sigma^2}{(2\rho ra\sin^{2} \theta)^2+[(r^{2}+a^{2})^2-a^2\Delta \sin^{2} \theta]\sin^{2} \theta(\Sigma-2\rho r)},
\label{motionphi}
\end{equation}
where $\mathcal{K}$ is the Carter constant.

The radial motion of photons is allowed only in the region where $\mathcal{R}(r)\ge 0$. Thus, the radial function $\mathcal{R}(r)$ plays a central role in determining the radial dynamics. We now focus on spherical photon orbits, namely null geodesics satisfying $r=\mathrm{const.}=r_{ph}$. For such orbits, the radial equation of motion in Eq. (\ref{motionr}) requires \cite{Chandrasekhar1998Themathematical,Teo2021Spherical}:
\begin{equation}
\mathcal{R}(r_{ph})=0, \quad \mathcal{R}^\prime (r_{ph}) =0, 
\label{rph1}
\end{equation}
where the prime denotes differentiation with respect to the radial coordinate $r$. On the other hand, from Eq. (\ref{motiontheta}), one has:
\begin{equation}
\Theta(\theta) \ge 0,\quad \theta \in [0,\pi ].
\label{rph2}
\end{equation}
Combining Eqs. (\ref{rph1}) and (\ref{rph2}) gives the condition that determines the photon region around the ANRBH:
\begin{equation}
16 r_{ph}^2 a^2 \sin^2 \theta (r_{ph}^{2}-2 \rho r_{ph}+a^{2}) \ge [4r_{ph}(r_{ph}^{2}+a^{2}-2 \rho r_{ph})+2(r_{ph}^{2}+a^{2} \cos ^{2} \theta)(\rho+r_{ph}\rho'-r_{ph}) ]^2.
\label{rph}
\end{equation}
Furthermore, the stability of spherical photon orbits can be determined by the sign of $\mathcal{R}^{\prime \prime}(r_{ph})$. The condition $\mathcal{R}^{\prime \prime}(r_{ph})<0$ corresponds to stable spherical photon orbits, whereas $\mathcal{R}^{\prime \prime}(r_{ph})>0$ corresponds to unstable spherical photon orbits. The boundary of the BH shadow is governed by unstable spherical photon orbits, which separate photon trajectories escaping to infinity from those captured by the BH.

Based on Eq. (\ref{rph}), together with the horizon condition $\Delta=0$ and the static limit surface condition $g_{tt}=0$, we plot the horizons, ergoregions, and photon regions of the ANRBH in Fig. \ref{figphoton}. The boundaries of the blue regions represent the BH horizons, while the gray hatched regions denote the ergoregions. Within the parameter ranges considered in this paper, the equation $\Delta=0$ generally yields two positive roots, corresponding to the Cauchy horizon and the event horizon, respectively. For a rotating BH, the region between the horizon and the static limit surface is referred to as the ergoregion. The orange regions represent unstable spherical photon orbits determined by $\mathcal{R}^{\prime \prime}(r_{ph})>0$, whereas the yellow regions correspond to stable spherical photon orbits with $\mathcal{R}^{\prime \prime}(r_{ph})<0$. As shown in Fig. \ref{figphoton}, the photon regions outside the event horizon correspond to unstable spherical photon orbits, while stable spherical photon orbit regions can appear inside the event horizon for specific parameter choices. (1) In the first row of Fig. \ref{figphoton}, we fix $a$, $Q$, and $\omega$ to investigate the influence of the global charge $K$ on the photon region. For the chosen parameter values, as $K$ decreases, the unstable photon region outside the event horizon gradually expands and becomes increasingly flattened. This indicates that the global charge $K$ has a pronounced influence on the external unstable photon region, which is directly related to the shadow boundary. (2) The second row of Fig. \ref{figphoton} shows the influence of the magnetic charge $Q$ on the BH geometry and photon region. As $Q$ increases, the separation between the inner and outer horizons decreases. Consequently, the blue region between the two horizons shrinks significantly. This suggests that the nonlinear electromagnetic field can substantially modify the horizon structure of the BH. Meanwhile, the distribution and shape of the photon region also change accordingly. In particular, when $Q=0$, the contribution from the nonlinear electromagnetic field vanishes, and a stable photon region determined by $\mathcal{R}^{\prime \prime}(r_{ph})<0$ appears inside the event horizon. (3) The third row of Fig. \ref{figphoton} shows that, when $a=0$, the rotating spacetime reduces to the spherically symmetric case. In this case, the ergoregion disappears, and the photon region reduces to the photon sphere. As the spin parameter increases, the separation between the static limit surface and the horizon becomes more pronounced, and the ergoregion expands significantly. For higher spin, stable spherical photon orbits may also appear inside the event horizon. (4) In addition, the second column of the figure shows that, when the other parameters are fixed, the photon region changes only slightly as $\omega$ increases. Overall, Fig. \ref{figphoton} illustrates the effects of the anisotropic matter field, the nonlinear electromagnetic field, and the spin parameter on the distribution of photon regions around the ANRBH.

\section{$\text{Shadow and Energy Emission Rate of the Anisotropic Nonlinear Magnetic Charged Rotating Black Hole}$}

In recent years, the EHT Collaboration has successively released images of the BHs M87* and Sgr A* using very-long-baseline interferometry \cite{Azulay2019FirstM87,Akiyama2019FirstM87,Akiyama2022FirstSagittariusA}. These images show a bright ring surrounding a central dark region, and this dark region is usually referred to as the BH shadow. Theoretically, the BH shadow can be regarded as the optical projection of the strong field spacetime geometry and is usually described by celestial coordinates on the observer's sky. Its apparent shape is closely related to the spacetime structure, critical photon orbits, and the observer's position. Therefore, BH shadows provide an important observational window for investigating BH properties in the strong gravity regime. Following the analysis of the photon regions of the ANRBH, we now investigate the shadow properties of this BH.

\subsection{Shadow}

The equations of photon motion and the condition for the photon region have been derived in the previous section. To construct the shadow cast seen by a distant observer, we introduce the impact parameters:
\begin{equation}
\xi=\dfrac{L_z}{E},
\end{equation}
\begin{equation}
\eta=\dfrac{\mathcal{K}}{{E}^{2}},
\end{equation}
where $\xi$ encodes the photon motion around the rotation axis, while $\eta$ characterizes its deviation from the equatorial plane \cite{Perlick2022Calculating}. For the ANRBH considered in this work, solving the spherical photon orbit conditions in Eq. (\ref{rph1}) gives the critical impact parameters $\xi$ and $\eta$, which can be written as functions of the radius $r_{ph}$ of spherical photon orbits and the physical parameters $a$, $\omega$, $Q$, and $K$:
\begin{equation}
\xi=\frac{\left(a^{2}-3 r_{ph}^{2}\right) \rho+r_{ph}\left(a^{2}+r_{ph}^{2}\right)\left(1+\rho^{\prime}\right)}{a\left[\rho+r_{ph}\left(-1+\rho^{\prime}\right)\right]},
\end{equation}
\begin{equation}
\eta=\frac{r_{ph}^{3}\left[2 \rho\left(2 a^{2}+3 r_{ph}^{2}+3\rho^{\prime} r_{ph}^{2} \right)-r_{ph}^{3}-9\rho^{2} r_{ph} -2\rho^{\prime}r_{ph}\left(2 a^{2} +r_{ph}^{2}\right) -{\rho^{\prime}}^{2}r_{ph}^{3} \right]}{a^{2}\left[\rho+r_{ph}\left(-1+\rho^{\prime}\right)\right]^{2}}.
\end{equation}
These critical impact parameters determine the critical photon trajectories that form the boundary of the BH shadow \cite{Jusufi2019Black,Yunusov2024Rotating}. To obtain the shadow cast on the observer's image plane, we relate these parameters to the corresponding celestial coordinates. We assume that the observer is located at $(r_o,\theta_o)$, with $r_o \to \infty$, where $\theta_o$ denotes the observer's polar angle. For a static observer at infinity, we introduce the celestial coordinates $(\alpha,\beta)$ as follows \cite{Bardeen1972Rotating}:
\begin{equation}
\alpha=\lim_{r_{ o} \to \infty} \left( -r_{ o}^2 \sin \theta_{ o} \frac{d\phi}{dr}\mid_{ (r_o,\theta_o)} \right),
\end{equation}
\begin{figure}[H]
\centering
\includegraphics[width=0.48\textwidth]{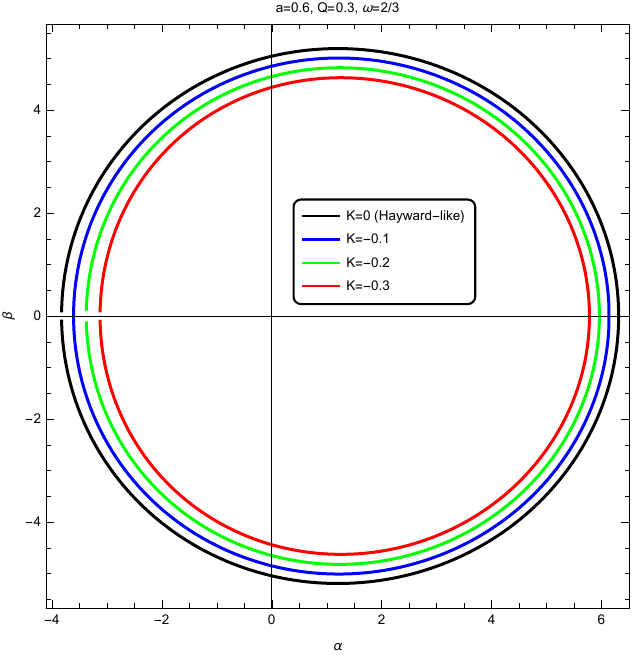}
\includegraphics[width=0.48\textwidth]{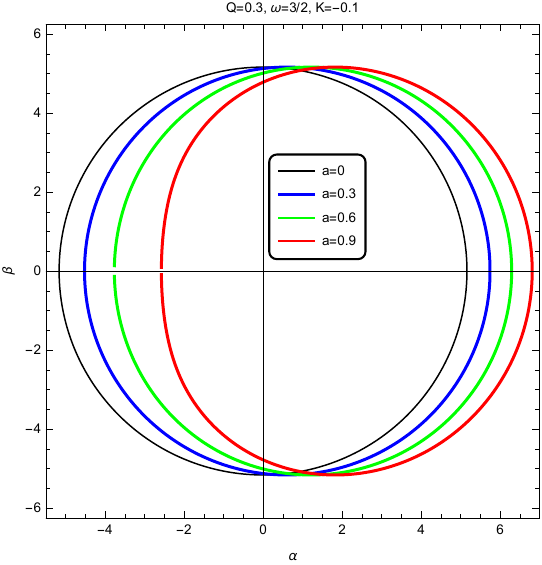}
\includegraphics[width=0.48\textwidth]{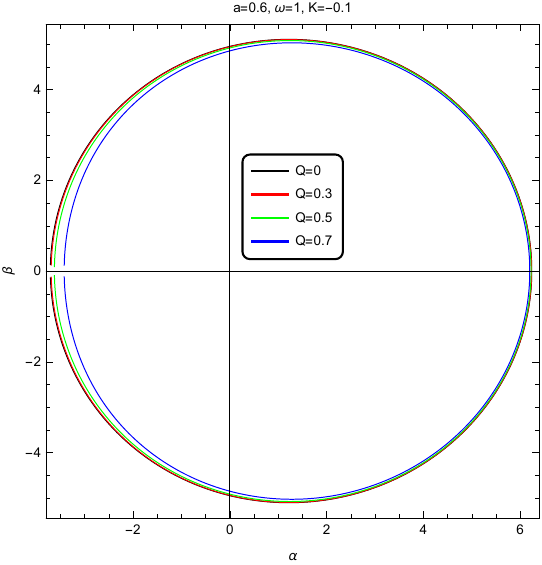}
\includegraphics[width=0.48\textwidth]{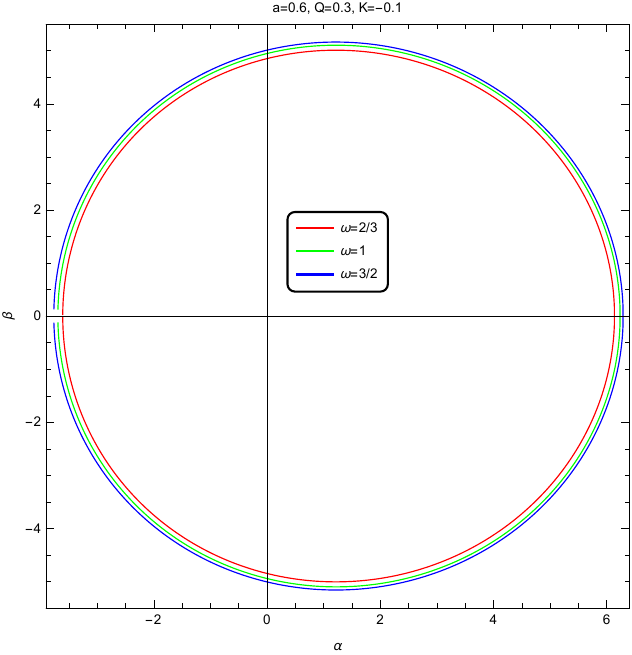}
\caption{Shadow casts of the anisotropic nonlinear magnetic charged rotating black hole.}
\label{figcast}
\end{figure}
\begin{equation}
\beta=\lim_{r_{ o} \to \infty} \left(r_o^2 \frac{d\theta}{dr}\mid_{ (r_o,\theta_o)}\right),
\end{equation}
where $\alpha$ and $\beta$ denote the horizontal and vertical coordinates on the observer's screen, respectively. From the photon equations of motion in Eqs. (\ref{motiont})-(\ref{motionphi}), the celestial coordinates $(\alpha,\beta)$ can be expressed in terms of the critical impact parameters $\xi$ and $\eta$ as:
\begin{equation}
\alpha=-\frac{\xi}{\sin \theta_o},
\end{equation}
\begin{equation}
\beta=\pm \sqrt{\eta +a^2 \cos^2\theta_o-\xi^2 \cot^2 \theta_o}.
\end{equation}
The above relations directly connect the critical photon orbits with the shadow cast seen by a distant observer \cite{Hou2018Rotating,Papnoi2022Rotating,Haroon2019Shadow}. For an equatorial observer with $\theta_o=\pi/2$, these relations reduce to:
\begin{equation}
\alpha=-\xi,
\end{equation}
\begin{equation}
\beta=\pm \sqrt{\eta}.
\end{equation}
For given BH parameters, the critical impact parameters $\xi$ and $\eta$ associated with unstable spherical photon orbits determine the shadow boundary in the $(\alpha,\beta)$ plane. The shadow casts of the ANRBH are constructed from these relations, and their dependence on the physical parameters is analyzed below.

Fig. \ref{figcast} presents the shadow casts of the ANRBH for an equatorial observer with varying physical parameters. To examine the effect of each physical parameter separately, we vary only one parameter in each panel while keeping the others fixed at the values indicated in Fig. \ref{figcast}. (1) The upper-left panel of Fig. \ref{figcast} shows that the BH shadow gradually shrinks as the global charge parameter $K$ decreases. Relative to the Hayward-like rotating case ($K=0$), decreasing $K$ reduces the overall size of the shadow, while the shadow cast remains approximately circular. This indicates that the anisotropic matter field mainly affects the shadow size rather than its shape. (2) The upper-right panel of Fig. \ref{figcast} illustrates how the spin parameter $a$ affects the shadow cast. For $a=0$, the BH reduces to the spherically symmetric case, and the shadow cast is circular. As $a$ increases, the shadow boundary gradually departs from circularity and becomes increasingly asymmetric. At higher spin, the shadow is shifted horizontally, and its left edge becomes more flattened. As a result, the cast develops a D-shaped profile resembling the Kerr BH shadow. This shows that the spin parameter mainly controls the degree of shadow distortion. (3) We next examine the influence of the nonlinear electromagnetic field parameter $Q$ on the shadow cast. As shown in the lower-left panel of Fig. \ref{figcast}, increasing $Q$ only weakly affects the shadow cast. Compared with the $Q=0$ case, increasing $Q$ slightly shifts the shadow boundary and produces only a small variation in the asymmetric deformation on the left side of the shadow. (4) The lower-right panel of Fig. \ref{figcast} describes the effect of the state parameter $\omega$ on the shadow cast. As $\omega$ increases, the shadow cast expands slightly outward, indicating that $\omega$ mainly influences the overall size of the shadow. Compared with $K$ and $a$, the parameters $Q$ and $\omega$ induce weaker changes in the shadow size and shape.

\begin{figure}[H]
\centering
\includegraphics[width=0.24\textwidth]{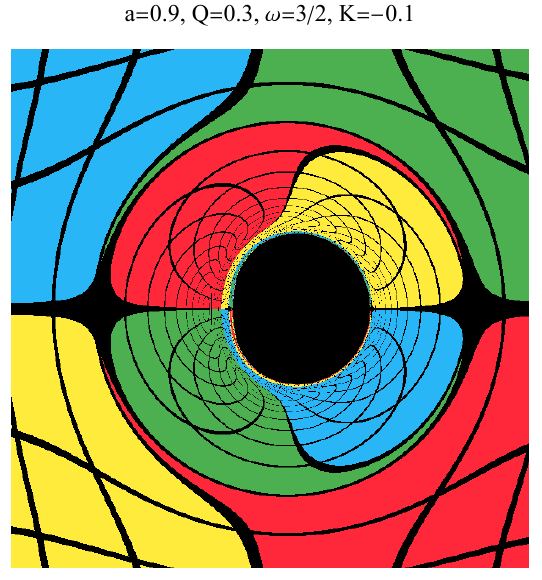}
\includegraphics[width=0.24\textwidth]{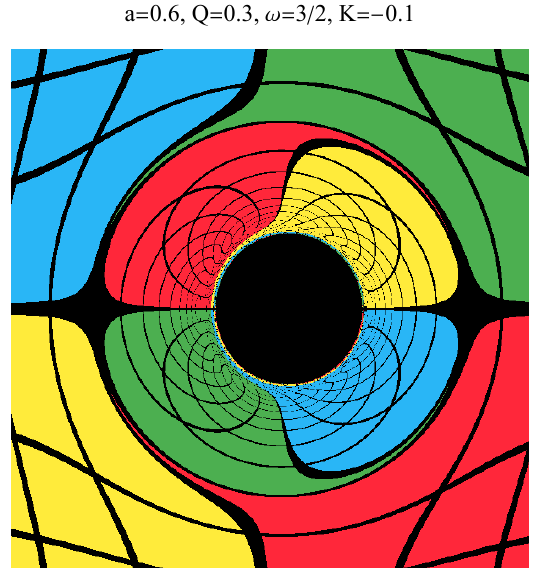}
\includegraphics[width=0.24\textwidth]{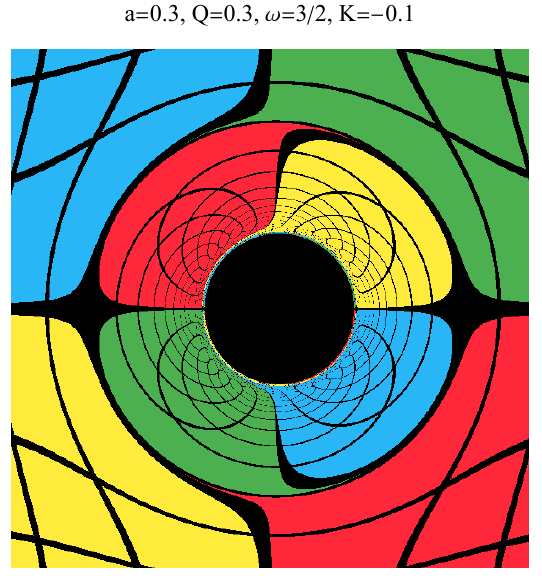}
\includegraphics[width=0.24\textwidth]{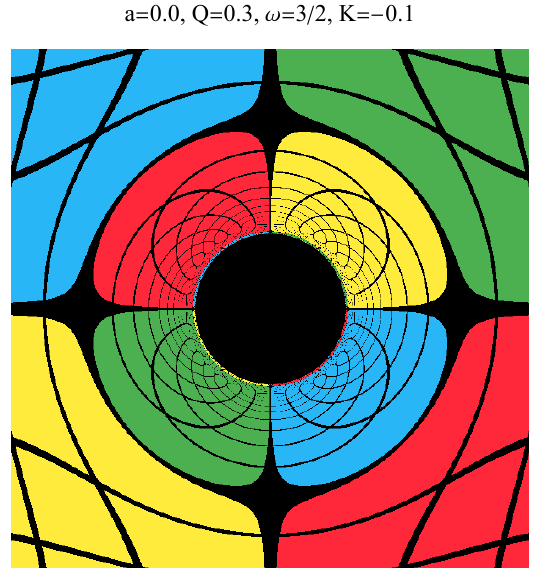}
\includegraphics[width=0.24\textwidth]{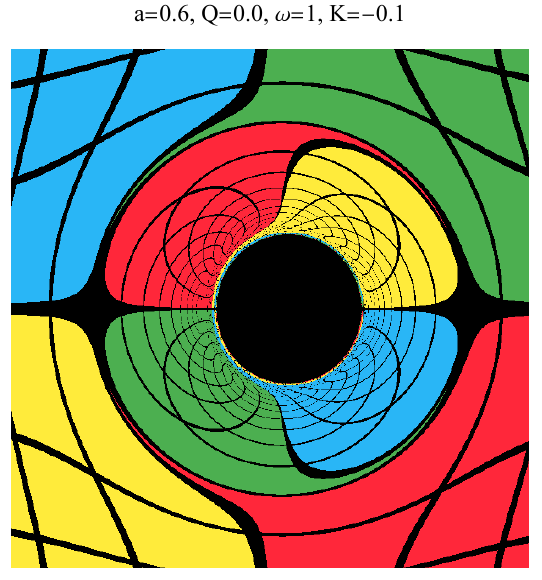}
\includegraphics[width=0.24\textwidth]{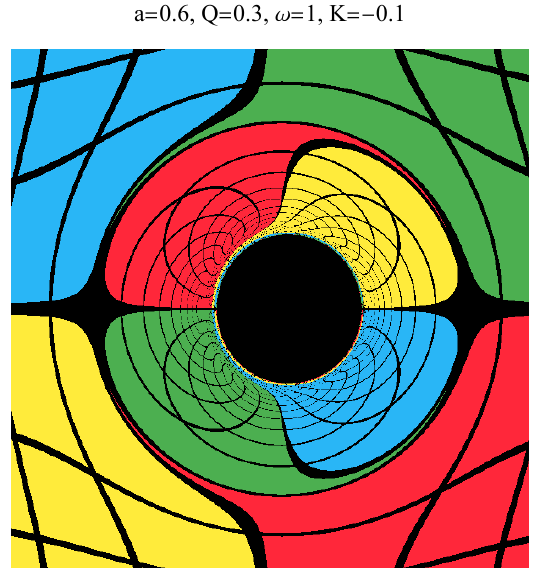}
\includegraphics[width=0.24\textwidth]{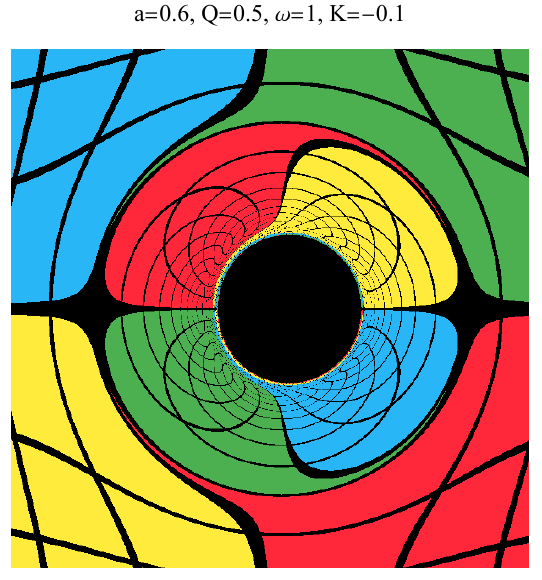}
\includegraphics[width=0.24\textwidth]{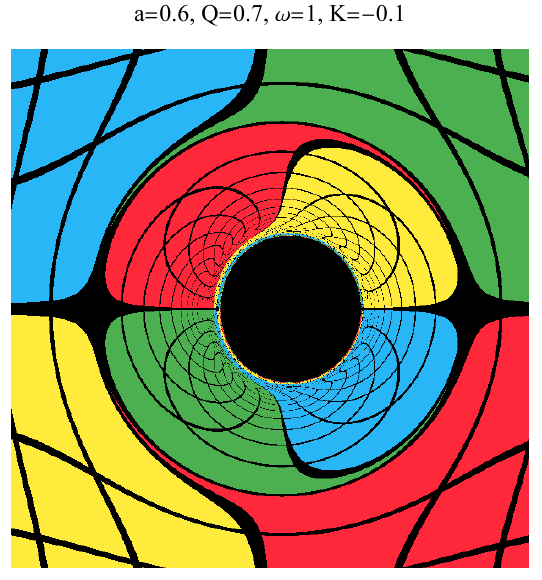}
\includegraphics[width=0.24\textwidth]{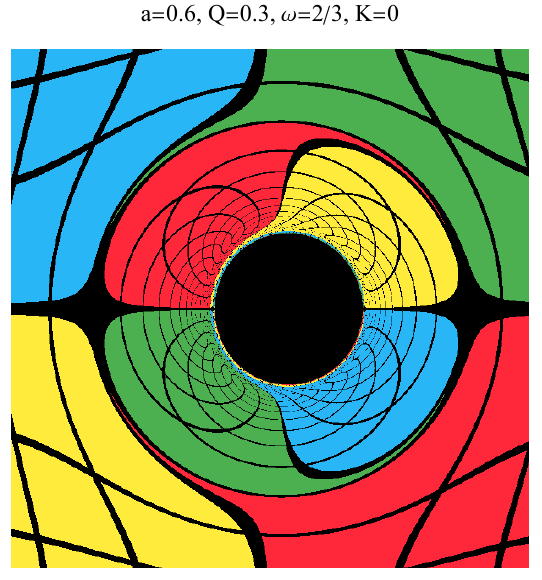}
\includegraphics[width=0.24\textwidth]{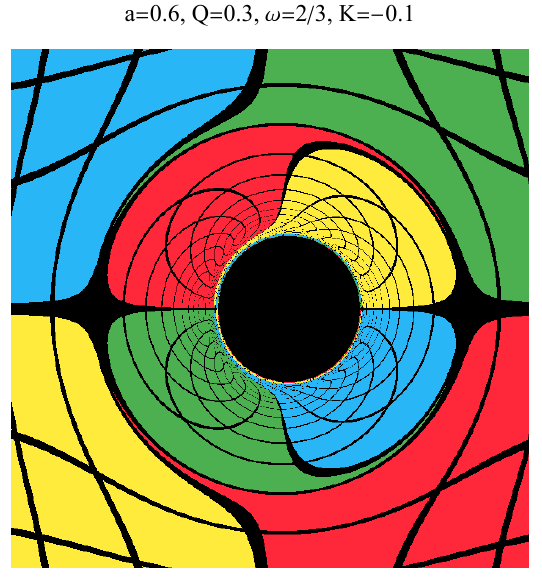}
\includegraphics[width=0.24\textwidth]{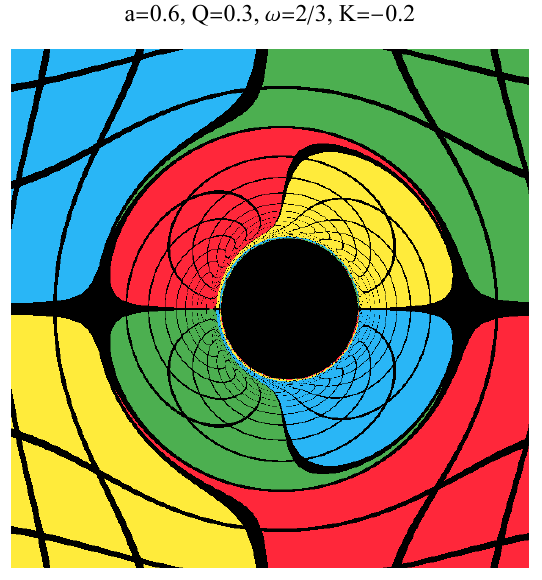}
\includegraphics[width=0.24\textwidth]{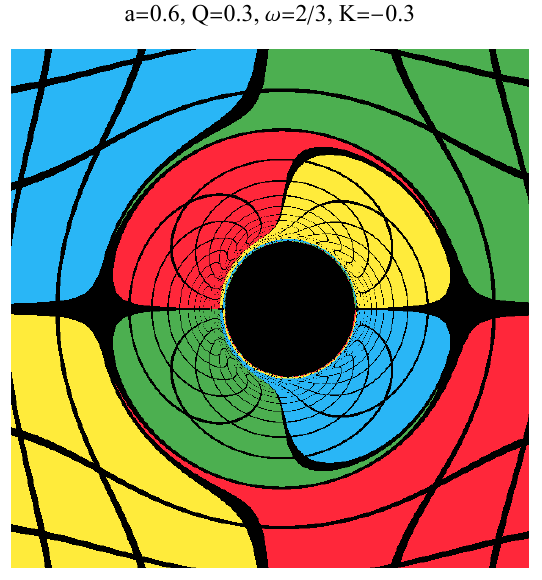}
\caption{Shadow images of the anisotropic nonlinear magnetic charged rotating black hole.}
\label{figshadow}
\end{figure}

To visualize the formation of the ANRBH shadow more intuitively, we construct BH shadow images using the backward ray-tracing method \cite{Liu2024Light,Hu2021QED}. In this setup, the distant celestial sphere is divided into four colored regions: red, yellow, blue, and green. A reference grid of latitude and longitude lines is also placed on the celestial sphere, with adjacent grid lines separated by an angular interval of $\pi/18$ ($10^\circ$). Specifically, in backward ray tracing, rays captured by the BH form the central dark shadow, whereas rays reaching the distant celestial sphere carry the corresponding background colors. These images show not only the shape and size of the central shadow, but also how the strong gravitational field of the BH affects light propagation, as reflected by the distortion of the background color regions and reference grid. Following the approach adopted in Refs. \cite{Zhong2021QED,Liu2025Shadow}, Fig. \ref{figshadow} presents the shadow images for different physical parameters as seen by an equatorial observer with $\theta_o=90^{\circ}$.

The first row of Fig. \ref{figshadow} shows a circular central dark shadow for $a=0$. The surrounding background grid is bent by the strong gravitational field of the BH, but the image as a whole remains symmetric. As the spin parameter $a$ increases, the shadow region gradually departs from circularity. Meanwhile, the background color boundaries and grid lines near the shadow become noticeably bent and tilted. This suggests that BH spin modifies the shadow shape and, through the frame-dragging effect, alters photon trajectories, thereby enhancing the asymmetric distortion of the image. The second and third rows of Fig. \ref{figshadow} show the effects of the nonlinear electromagnetic field parameter $Q$ and the global charge parameter $K$, respectively. The second column displays the shadow images for different values of the state parameter $\omega$. The effects of $Q$ and $\omega$ are relatively weak, whereas decreasing $K$ leads to a significant contraction of the central shadow region. Overall, Fig. \ref{figshadow} directly visualizes how the strong gravitational field of the BH bends and distorts the distant celestial sphere. The images show that BH spin mainly controls the asymmetric distortion of the shadow, while the global charge parameter $K$ has the most pronounced effect on the shadow size. By contrast, the nonlinear electromagnetic field parameter $Q$ and the state parameter $\omega$ have relatively weak effects. These numerical and imaging results provide a clear visual basis for understanding how rotation and different physical parameters affect the BH shadow.

\subsection{Shadow Observables}

The shape and size of a BH shadow provide important information for testing gravity in the strong field regime. Having obtained the shadow casts and ray-tracing images of the ANRBH, we now introduce a set of observables to quantify how different physical parameters affect the shadow size and deformation.

Hioki and Maeda introduced the shadow radius $R_s$ and the distortion parameter $\delta_s$ to characterize the overall size of a rotating BH shadow and its deviation from circularity \cite{Hioki2009Measurement}. This construction provides an intuitive parametrization of rotating BH shadows and has been widely used in studies of BH shadows in various spacetimes \cite{Fu2022Shadow,Liu2025Lorentz}. To define these quantities, three characteristic points are first selected on the shadow boundary: the top point $(\alpha_t,\beta_t)$, the bottom point $(\alpha_b,\beta_b)$, and the rightmost point $(\alpha_r,0)$. These three points uniquely determine a reference circle, whose radius is defined as the shadow radius $R_s$:
\begin{eqnarray}
R_s = \frac{(\alpha_t-\alpha_r)^2+\beta_t^2}{2|\alpha_r-\alpha_t|}.
\label{Rs}
\end{eqnarray}
For a perfectly circular shadow, the above expression reduces to the radius of the corresponding circle. For a rotating BH, the shadow is generally displaced and distorted, so $R_s$ provides a measure of its overall size. To quantify the deviation of the shadow from the reference circle, one introduces the distortion parameter $\delta_s$. Here, $(\alpha_l,0)$ denotes the leftmost intersection of the shadow cast with the horizontal axis $\beta=0$, while $(\alpha_l^{\prime},0)$ is the corresponding intersection point of the reference circle with the same axis. The distortion parameter is defined as:
\begin{eqnarray}
\delta_s = \frac{|\alpha_l'-\alpha_l|}{R_s},
\label{deltas}
\end{eqnarray}
where $|\alpha_l' - \alpha_l|$ measures the horizontal displacement of the left edge of the shadow relative to the reference circle. Thus, $\delta_s$ quantifies the deviation of the shadow from a perfect circle.

Kumar and Ghosh \cite{Kumar2020Black} further introduced the shadow area $A$ and the oblateness $D$ as observables for characterizing the shadow size and its deviation from circularity. These quantities are defined by:
\begin{eqnarray}
A = 2\int_{r_{ph}^{-}}^{r_{ph}^+}\left( \beta(r_{ph}) \frac{d\alpha(r_{ph})}{dr_{ph}}\right)dr_{ph},
\label{A}
\end{eqnarray}
\begin{eqnarray}
D = \frac{\alpha_r-\alpha_l}{\beta_t-\beta_b},
\label{D}
\end{eqnarray}
where $r_{ph}^{\pm}$ denote the radii of the prograde and retrograde unstable spherical photon orbits. These orbits determine the shadow boundary on the observer's image plane. For a spherically symmetric spacetime, $D=1$. A deviation of $D$ from unity indicates that the shadow has different horizontal and vertical extents, thereby quantifying the shape distortion caused by rotation or additional physical parameters \cite{Tsupko2017Analytical}.

\begin{figure}[H]
\centering
\includegraphics[width=0.49\textwidth]{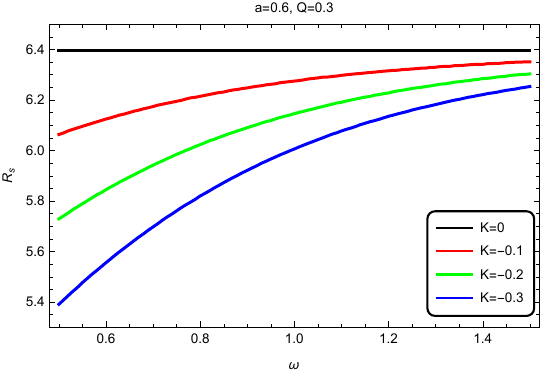}
\includegraphics[width=0.49\textwidth]{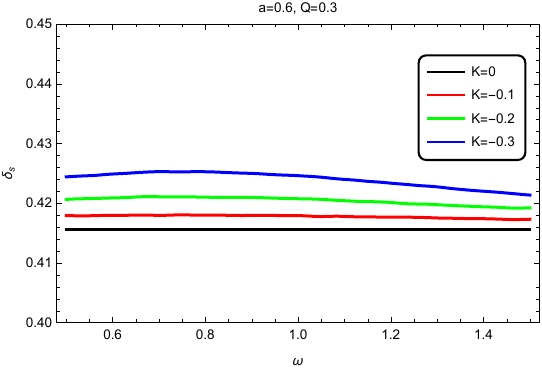}
\includegraphics[width=0.49\textwidth]{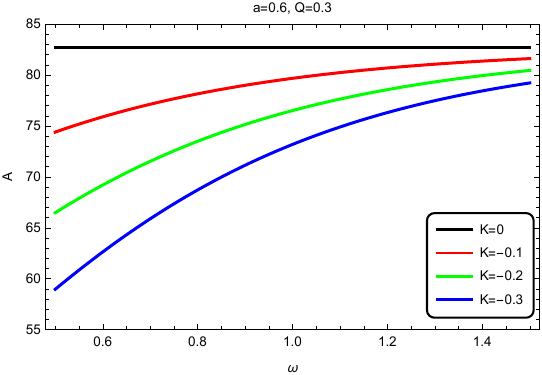}
\includegraphics[width=0.49\textwidth]{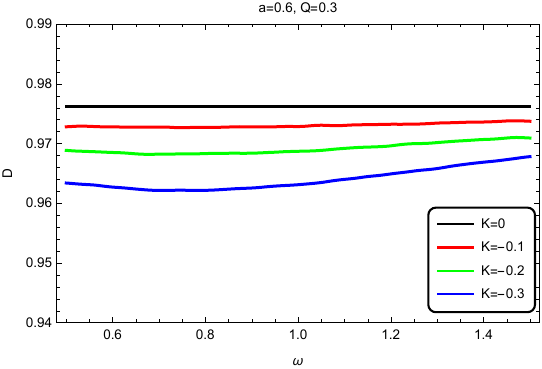}
\caption{Shadow observables of the anisotropic nonlinear magnetic charged rotating black hole. The panels show the variations of the shadow radius, distortion parameter, shadow area, and oblateness with the anisotropic matter field parameters.}
\label{figRsD}
\end{figure}

We now focus on the influence of the anisotropic matter field on the shadow observables of the ANRBH. Using Eqs. (\ref{Rs})-(\ref{D}), we calculate the shadow radius $R_s$, distortion parameter $\delta_s$, shadow area $A$, and oblateness $D$ for fixed $a=0.6$ and $Q=0.3$. Fig. \ref{figRsD} shows these observables as functions of the state parameter $\omega$, with different curves corresponding to different values of the global charge parameter $K$. (1) The upper-left panel of Fig. \ref{figRsD} shows the behavior of the shadow radius $R_s$. When $K=0$, the anisotropic matter contribution vanishes, and $R_s$ is independent of $\omega$. For $K<0$, $R_s$ grows with $\omega$ and gradually approaches the $K=0$ curve. At fixed $\omega$, a decrease in $K$ leads to a pronounced reduction in $R_s$. This indicates that the global charge parameter tends to reduce the overall size of the shadow, while a larger $\omega$ weakens this reduction. (2) The distortion parameter $\delta_s$, shown in the upper-right panel of Fig. \ref{figRsD}, depends only weakly on $\omega$, and its overall variation remains small. Compared with the $K=0$ case, smaller values of $K$ lead to a slight increase in $\delta_s$, indicating a stronger deviation from circularity. This suggests that the anisotropic matter field changes the shadow size and slightly enhances its deviation from the reference circle. (3) The lower-left panel of Fig. \ref{figRsD} presents the behavior of the shadow area $A$. Since the area directly measures the overall size of the shadow, its behavior closely follows that of the shadow radius $R_s$. Specifically, $A$ increases as $\omega$ increases, while it decreases significantly as $K$ decreases. This further indicates that the anisotropic matter field affects the shadow geometry mainly by changing the shadow size. (4) The lower-right panel of Fig. \ref{figRsD} shows that the oblateness $D$ changes only slightly with $\omega$, but decreases as $K$ decreases. This indicates that smaller $K$ increases the deviation of the shadow from circularity and enhances its shape distortion. These results show that, within the parameter range considered here, the anisotropic matter field affects the shadow size more strongly than the shadow shape. Specifically, as the global charge parameter $K$ decreases, both the shadow radius $R_s$ and the shadow area $A$ decrease significantly, whereas increasing the state parameter $\omega$ leads to a noticeable increase in $R_s$ and $A$. By contrast, the distortion parameter $\delta_s$ and the oblateness $D$ change only slightly as the anisotropic matter field parameters vary. Therefore, the influence of the anisotropic matter field on the shadow observables is mainly reflected in the variation of the shadow size, with a weaker influence on the shape distortion.

\subsection{Energy Emission Rate}

We now investigate the energy emission rate of the ANRBH and examine the effects of the anisotropic matter and nonlinear electromagnetic parameters. In addition to the shadow and its observables, the energy emission rate provides another way to characterize the physical properties of the BH from the perspective of Hawking radiation. The BH shadow is closely associated with the photon capture region determined by unstable photon orbits. In the geometric-optics limit, the high-energy absorption cross section $\sigma_{lim}$ can be approximated by the effective area of this capture region \cite{Decanini2011Universality,Sanchez1978Elastic}. Therefore, the shadow radius not only measures the size of the BH shadow, but also provides an estimate of the effective absorption cross section in the high-energy limit. From a physical point of view, Hawking radiation indicates that BHs can emit radiation with a thermal spectrum \cite{Hawking1975Particle}. The corresponding energy emission rate is governed by both the absorption cross section and the Hawking temperature. Thus, analyzing the energy emission rate helps to connect the geometric features of the BH shadow with its thermal radiation behavior.

For a spherically symmetric BH, $\sigma_{lim}$ is associated with the geometric capture area determined by the photon sphere. In the rotating case, this quantity is often estimated using an effective shadow radius. Following Ref. \cite{Wei2013Observing}, the limiting absorption cross section is approximated as:
\begin{equation}
\sigma_{lim}\simeq \pi R_s^2 ,
\end{equation}
where $R_s$ is the shadow radius defined in Eq. (\ref{Rs}). Accordingly, the energy emission rate per unit frequency and unit time can be written as:
\begin{equation}
\frac{d^2E(\varpi)}{d\varpi dt}=\frac{2\pi^2\sigma_{lim}}{e^{\frac{\varpi}{T_H}}-1}\varpi^3,
\label{ERsTH}
\end{equation}
where $\varpi$ denotes the emission frequency, and $T_H$ is the Hawking temperature of the BH. This expression indicates that the energy emission rate is affected by two key factors: the effective absorption cross section estimated from the shadow radius and the thermal factor governed by the Hawking temperature. For the anisotropic nonlinear magnetic charged rotating BH considered here, the Hawking temperature is related to the surface gravity at the event horizon and takes the form:
\begin{equation}
T_H=\frac{r_H^{2 \omega}\left[-2 Q^{3} r_H^{2}+r_H^{5}-a^{2}\left(4 Q^{3}+r_H^{3}\right)\right]+K\left[2 Q^{3} r_H^{2}(1+\omega)+r_H^{5}(-1+2 \omega)\right]}{4 \pi r_H^{1+2 \omega} \left(a^{2}+r_H^{2}\right)\left(Q^{3}+r_H^{3}\right)},
\end{equation}
here, $r_H$ denotes the event horizon radius determined by $\Delta(r_H)=0$. The above expressions show that the parameters $a$, $Q$, $\omega$, and $K$ influence the energy emission rate in two ways. First, they affect the shadow radius $R_s$, thereby changing the effective absorption cross section in the high-energy limit. Second, they shift the event horizon radius $r_H$, which changes the Hawking temperature $T_H$. In this sense, the energy emission rate encodes both the geometric information of the BH shadow and the thermodynamic properties of the horizon. It can therefore be used to analyze how these physical parameters affect the radiation behavior of the BH.

\begin{figure}[H]
\centering
\includegraphics[width=0.32\textwidth]{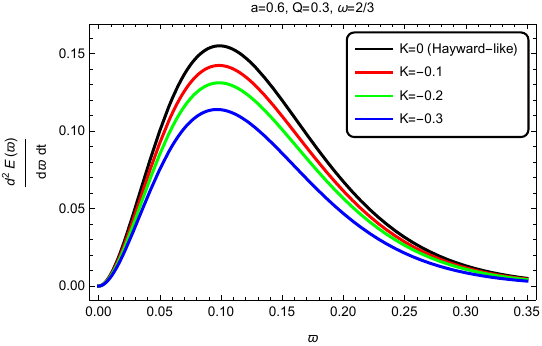}
\includegraphics[width=0.32\textwidth]{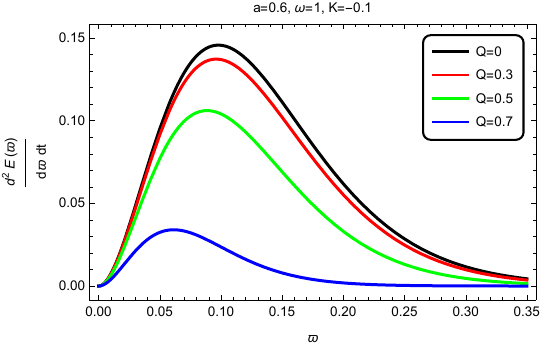}
\includegraphics[width=0.32\textwidth]{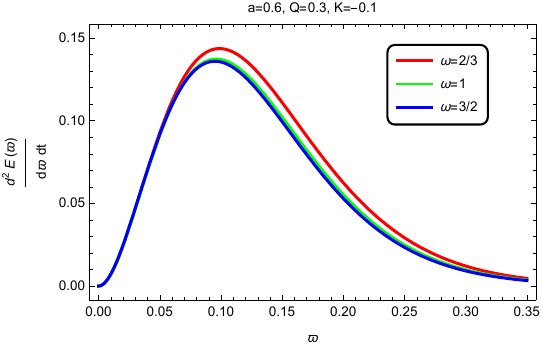}
\caption{Variation of the energy emission rate of the anisotropic nonlinear magnetic charged rotating black hole with frequency.}
\label{figERsTH}
\end{figure}

Using Eq. (\ref{ERsTH}), we plot the energy emission rate of the ANRBH as a function of the emission frequency $\varpi$ in Fig. \ref{figERsTH}. For the parameter values considered here, all curves exhibit a similar nonmonotonic spectral behavior. The emission rate rises rapidly from zero in the low-frequency region, reaches a maximum at a characteristic frequency, and then gradually decreases at higher frequencies. (1) As shown in the left panel of Fig. \ref{figERsTH}, for fixed $a=0.6$, $Q=0.3$, and $\omega=2/3$, decreasing the global charge parameter $K$ lowers the emission peak, indicating that smaller values of $K$ weaken the energy emission in the chosen parameter range. (2) The middle panel illustrates the role of the nonlinear electromagnetic parameter $Q$. As $Q$ increases, the energy emission rate decreases markedly. For larger values of $Q$, the emission peak is further suppressed and shifts toward lower frequencies. This indicates that the nonlinear electromagnetic parameter has a strong weakening effect on the energy emission rate. (3) In contrast, the right panel shows that the state parameter $\omega$ has a relatively weak influence. With the other parameters fixed, varying $\omega$ produces only small changes in the spectrum. Increasing $\omega$ slightly lowers the peak and mildly changes the high-frequency decay region.

\section{$\text{Conclusion}$}

In this paper, we investigated the shadow properties of a rotating BH in the presence of both a nonlinear electromagnetic field and an anisotropic matter field. The analysis includes photon regions, shadow structure, shadow observables, and the energy emission rate. In Sec. II, we reviewed the basic properties of the anisotropic nonlinear magnetic charged rotating BH solution. The nonlinear electromagnetic field is characterized by the magnetic charge parameter $Q$, whereas the anisotropic matter field is described by the global charge parameter $K$ and the state parameter $\omega$. These parameters enter the metric function and thereby affect the event horizon, photon motion, and shadow characteristics of the BH.

First, we analyzed the horizon structure, ergoregion distribution, photon motion, and photon regions around the ANRBH. Using the Hamilton--Jacobi equation, we derived the photon equations of motion, where the radial and angular parts are separated by the Carter constant. We then obtained the condition for the photon region from the spherical photon orbit conditions and the angular constraint. It is used to identify the allowed photon regions around the ANRBH and to examine their stability. The photon-region analysis shows that the physical parameters can strongly affect the distribution of spherical photon orbits around the BH. For the parameter ranges considered here, the photon regions outside the event horizon are composed of unstable spherical photon orbits, whereas stable spherical photon orbits may appear inside the horizon for certain parameter choices. The influence of the global charge $K$ is relatively pronounced. As $K$ decreases, the outer unstable photon region expands and becomes slightly flattened, while an inner photon region gradually appears and grows. Increasing the magnetic charge $Q$ mainly changes the locations of the BH horizons. By contrast, increasing the state parameter $\omega$ only weakly affects the overall structure of the photon region. The spin parameter $a$, however, plays a more important role. As $a$ increases, the ergoregion expands and the photon region becomes increasingly deformed.

Next, we introduced the impact parameters and constructed the BH shadow from the critical photon orbits using the celestial coordinates. The corresponding shadow images were further generated by the backward ray-tracing method. The results show that the global charge parameter $K$ mainly changes the overall shadow size. As $K$ decreases, the shadow gradually shrinks, indicating that the anisotropic matter field tends to reduce the shadow size. The spin parameter $a$ mainly affects the asymmetric deformation of the shadow. With increasing $a$, the shadow boundary deviates from a circular profile and develops a D-shaped structure similar to that of the Kerr BH. By contrast, the nonlinear electromagnetic parameter $Q$ and the state parameter $\omega$ have relatively weak effects on the shadow cast, producing only small changes in the shadow boundary and size. To further examine the influence of the anisotropic matter field on the geometrical characteristics of the shadow, we calculated four observables: the shadow radius $R_s$, distortion parameter $\delta_s$, shadow area $A$, and oblateness $D$. These quantities characterize the size and deformation of the shadow. We found that decreasing $K$ significantly reduces the shadow radius $R_s$ and shadow area $A$, while the distortion parameter $\delta_s$ and oblateness $D$ change only slightly. In addition, larger values of $\omega$ lead to a gradual increase in $R_s$ and $A$, suggesting that increasing $\omega$ can partly offset the reduction in the shadow size caused by decreasing $K$. These results indicate that, within the parameter range considered here, the anisotropic matter field has a more significant effect on the shadow size than on the shadow shape.

Finally, we analyzed the energy emission rate of the ANRBH. Using the geometric-optics approximation, we estimated the high-energy absorption cross section from the shadow radius. In this way, the energy emission rate can be expressed in terms of the shadow radius $R_s$ and the Hawking temperature $T_H$. The results show a typical thermal radiation profile: as the emission frequency increases, the energy emission rate starts from zero in the low-frequency region, reaches a maximum at a characteristic frequency, and then gradually decreases at higher frequencies. The graphical analysis shows that the global charge parameter $K$ and the nonlinear electromagnetic parameter $Q$ have relatively pronounced effects on the radiation spectrum. As $K$ decreases, the peak of the energy emission rate gradually decreases. As $Q$ increases, the emission rate is markedly suppressed, and the peak shifts toward lower frequencies. By contrast, the state parameter $\omega$ has only a weak influence. These results show that the energy emission rate depends on both the high-energy absorption cross section and the Hawking temperature. The physical parameters affect the absorption cross section through the shadow radius and change $T_H$ through the horizon structure. In this way, they determine the peak height and peak position of the energy-emission spectrum. In summary, the nonlinear electromagnetic field and the anisotropic matter field affect the photon motion, shadow structure, and energy emission behavior of the ANRBH in different ways. The present results therefore provide a theoretical basis for using BH shadows to probe the possible effects of anisotropic matter fields.

\textbf{\ Acknowledgments }
The research work is supported by the National Natural Science Foundation of China (12175095), and supported by  LiaoNing Revitalization Talents Program (XLYC2007047).

\end{document}